\documentclass[12pt]{article}\usepackage{qInfoStyle}

\begin{document}

\title{Quantum Entanglement and \\
Conditional Information Transmission}

\author{Robert R. Tucci\\
        P.O. Box 226\\ 
        Bedford,  MA   01730\\
        tucci@ar-tiste.com}

\date{ \today} 

\maketitle

\vskip2cm
\section*{Abstract}
We propose a new measure of quantum entanglement. Our measure is defined in terms of
conditional
information transmission for a Quantum Bayesian Net. We show that our measure 
is identically equal  to the Entanglement of Formation in the case of a 
bipartite (two listener) system occupying a pure state. 
In the case of mixed states, the relationship between these 
two measures is not known yet. 
We discuss some properties of our measure.
Our measure can be easily and naturally
generalized to handle $n$-partite ($n$-listener) systems. 
It is non-negative for any $n$.
It vanishes for conditionally separable states with $n$ listeners.
It is symmetric under permutations of the $n$ listeners. It decreases
if listeners are merged, pruned or removed. Most promising of all,
it is intimately connected with the Data Processing Inequalities.
We also find a new upper bound for classical 
mutual information which is  of interest in its own right.

\newpage

\BeginSection{Introduction}\label{sec:intro}
Quantum entanglement is at the very heart of Quantum Mechanics so
there is a vast amount of literature on the subject.
Of particular interest to workers
in the field of Quantum Information Theory are the issues of 
quantification and manipulation of entanglement.
An important step in that direction was taken
in Refs.\cite{NoHootters}-\cite{MonsterHootters}.
These references introduced measures of entanglement called  entanglement of  
formation and of distillation. Since Refs.\cite{NoHootters}-\cite{MonsterHootters}, the implications of
these two measures have been explored and clarified considerably by many workers\cite{mixture}.
And yet, the quantification of entanglement for mixed states and  for more than two listeners
is still not well understood. 

The goal of this paper is to 
shed some light on the quantification of entanglement
by approaching it from a new perspective, that of Quantum Bayesian Nets
and conditional information transmission. For a review of Quantum Information
Theory from the point of view of quantum Bayesian nets, see Ref.\cite{TucciQInfo}.
Henceforth, we will assume that the reader
is familiar with the notation of Ref.\cite{TucciQInfo}.

\begin{figure}[h]
	\begin{center}
	\epsfig{file=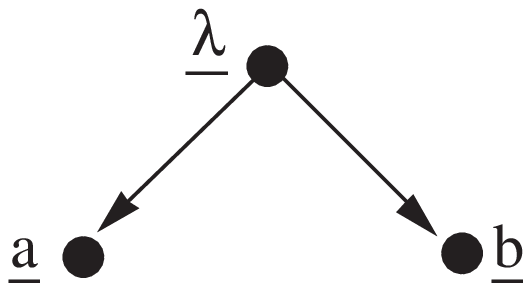}
	\caption{CB net in which $\rva$ and $\rvb$ are conditionally independent.}
	\label{fig:hidden-var}
	\end{center}
\end{figure}

For motivation, consider the CB net of Fig.\ref{fig:hidden-var}. This net satisfies

\beq
P(a,b, \lam) = P(a | \lam) P(b | \lam) P(\lam)
\;.
\label{eq:cond-indep-no-sum}\eeq
Summing the last equation over $\lam$, one gets 

\beq
P(a,b) = \sum_{\lam} P(a | \lam) P(b | \lam) P(\lam)
\;.
\label{eq:cond-indep-sum}\eeq
One says that $\rva$ and $\rvb$ are conditionally independent.
Eq.(\ref{eq:cond-indep-sum})
is often used as the starting point in the derivation of Bell Inequalities\cite{QFog}.
In that context, $\lam$ represents the hidden variables. 
As shown in Ref.\cite{TucciQInfo}, Eq.(\ref{eq:cond-indep-no-sum}) implies

\beq
H((\rva : \rvb) | \rvlam) = 0
\;.
\eeq
As we shall see in what follows, $S_\rho((\rva : \rvb) | \rvlam)$, 
the quantum mechanical counterpart of $H((\rva : \rvb) | \rvlam)$,
is NOT generally zero for a QB net with the graph of Fig.\ref{fig:hidden-var}. 
Thus, $S_\rho((\rva : \rvb) | \rvlam)$
appears to be a good measure of quantum entanglement, which is a 
phenomenon that does not occur classically. This paper is devoted to
discussing $S_\rho((\rva : \rvb) | \rvlam)$ and its generalizations.

\BeginSection{Entanglement of Formation}\label{sec:ent-of-for}

In this section, we will give a very brief review of the most 
basic aspects of the Entanglement of Formation.

Consider two Hilbert spaces $\hil_\rvx$ and $\hil_\rvy$ which need not
have the same dimension. Without loss of generality, we will assume 
that the dimension $N_\rvx$ of $\hil_\rvx$ is less than or equal to
the dimension $N_\rvy$ of $\hil_\rvy$ 

The {\it entanglement of formation $E_F$ for a bipartite
pure state} $\ket{\psi}\in \hil_\rvx \otimes \hil_\rvy$
is defined by 

\beq
E_F( \ket{\psi} ) = 
S[ \tr_\rvy (\ket{\psi}\bra{\psi})]
\;.
\label{eq:ef-pure}\eeq

Consider any density matrix $\rho$. If $\cale = \{  (w_a, \ket{\psi_a}) | \forall a\}$ satisfies  

\beq
\rho = \sum_a w_a \ket{\psi_a}\bra{\psi_a}
\;,
\eeq
then we say $\cale$ is a  $\rho$-ensemble. (This
clearly defines an equivalence relationship).
Ref.\cite{Hootters93} characterizes 
all $\cale$ belonging to a given $\rho$. The {\it entanglement of formation $E_F$
for a bipartite mixed state} with 
density matrix $\rho$ acting on $\hil_\rvx \otimes \hil_\rvy$
is defined by

\beq
E_F(\rho) = \min_{\cale} \left\{ \sum_a w_a E_F(\ket{\psi_a}) \right\}
\;,
\label{eq:ef-mix}\eeq
where the minimum is taken over all ensembles 
$\cale = \{  (w_a, \ket{\psi_a}) | \forall a\}$ which are $\rho$-ensembles.

First, let us  consider
$E_F$ for pure states. Let $\psi$ be the rectangular matrix 
with entries $\psi_{xy} = \av{x,y|\psi}$. We will often
denote $E_F( \ket{\psi} )$ by $E_F( \psi)$ or $E_F( \psi_{xy})$.
Thus,

\beq
E_F( \psi_{xy}) = S(\psi\psi^\dagger)
\;.
\label{eq:ef-pure-sum}\eeq
There always exist unitary matrices $U$ and $V$ such the 

\beq
U\psi V^\dagger = \markpsi
\;,
\label{eq:sing-value}\eeq
where the rectangular matrix $\markpsi$ is ``diagonal", in the sense that $\markpsi_{xy}=0$ if $x\neq y$. Eq.(\ref{eq:sing-value})
is called the Singular Value Decomposition\cite{Noble} of $\psi$.
Define $p_i$ for $0 \leq i \leq N_\rvx -1$ by

\beq
U \psi \psi^\dagger U^\dagger =
\markpsi\markpsi^\dagger
=
diag( p_0, p_1 ,\ldots, p_{N_\rvx -1})
\;.
\label{eq:px-def}\eeq
Since $\av{\alpha | \markpsi\markpsi^\dagger| \alpha} \geq 0$
for any $\ket{\alpha}\in \hil_\rvx$, the $p_i$'s are non-negative
numbers. Furthermore, since 
$\tr( U \psi \psi^\dagger U^\dagger ) = \sum_{x,y} |\psi_{xy}|^2 = 1$, the 
$p_i$'s add up to one. Note that

\beq
\ket{\psi} = \sum_{x,y} \psi_{xy} \ket{x,y}
\;,
\eeq

\beq
\ket{\markpsi} = \sum_{x} \sqrt{p_x} \ket{\rvx=x,\rvy=x}
\;.
\label{eq:schmidt}\eeq
Eq.(\ref{eq:schmidt}) is called the Schmidt Representation\cite{Hootters93} of 
$\ket{\psi}$. It follows directly from the Singular Value Decomposition of $\psi$.
By Eq.(\ref{eq:ef-pure-sum}) and (\ref{eq:px-def}),

\beq
E_F(\psi_{xy}) = E_F(\markpsi_{x,y}) = -\sum_x p_x \log_2 p_x
\;.
\label{eq:ef-px}\eeq

\begin{figure}[h]
	\begin{center}
	\epsfig{file=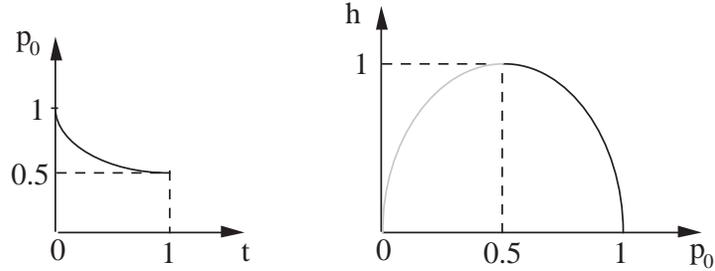, height=1.5in}
	\caption{Plot of functions $p_0(t)$ and $h(p_0)$.}
	\label{fig:po-h-plot}
	\end{center}
\end{figure}

For the remainder of this section, we will restrict our attention to 
the special case where $\rvx$ and $\rvy$ have just two states, 0 and 1.
In this case, $E_F(\psi_{xy}) = h(p_0)$, where $h$ is the binary
entropy function, and where 
$p_0$ and $p_1 = 1- p_0$ are the
eigenvalues of $\psi\psi^\dagger$.
Define complex numbers $K_0, K_1$ and $K$ by

\beq
\psi\psi^\dagger =
\left[ 
\begin{array}{cc}
K_0 & K \\
K^* & K_1
\end{array}
\right]
\;.
\eeq
Thus,

\begin{subequations}
\beq
K_0 = |\psi_{00}|^2 + |\psi_{01}|^2 
\;,
\eeq

\beq
K_1 = |\psi_{10}|^2 + |\psi_{11}|^2 
\;,
\eeq

\beq
K = \psi_{00} \psi^*_{10} + \psi_{01}\psi^*_{11} 
\;.
\eeq
\end{subequations}
The two eigenvalues of $\psi\psi^\dagger$ are

\begin{subequations}
\label{eq:p0-t-def}
\beq
p_0 =
\frac{ 1 + \sqrt{1-t}}{2}
\;,
\;\;\;
p_1 = 1-p_0
\;,
\eeq
where

\beq
t = 4 ( K_0 K_1 - |K|^2 ) = 4 | \psi_{00} \psi_{11} - \psi_{01}\psi_{10} |^2
\;.
\label{eq:t-def}\eeq
\end{subequations}
The {\it Bell Basis} is defined by 

\beq
\ket{B_f} = \frac{i^{f_0 + f_1}}{\sqrt{2}} 
(\ket{0, f_0} + (-1)^{f_1} \ket{1, \bar{f_0}})
\;,
\label{eq:bell-basis}\eeq
for $f=(f_0, f_1)\in Bool^2$. ($\bar{0}=1$  and $\bar{1}=0$.) If $x, y \in Bool$, then

\beq
\av{x, y | B_f} = \frac{i^{f_0 + f_1}}{\sqrt{2}} 
(\delta^{x,y}_{0, f_0} + (-1)^{f_1} \delta^{x,y}_{1, \bar{f_0}})
\;.
\eeq
Let $\alpha_j$ for $j\in Z_{0,3}$ be the components of $\ket{\psi}$ in the Bell Basis:

\beq
\ket{\psi} = 
\alpha_0 \ket{B_{00}}
+ \alpha_1 \ket{B_{01}}
+ \alpha_2 \ket{B_{10}}
+ \alpha_3 \ket{B_{11}}
\;.
\eeq
Then 

\beq
\psi_{00} = \frac{1}{\sqrt{2}}( \alpha_0 + i \alpha_1 )
\;,
\eeq

\beq
\psi_{01} = \frac{1}{\sqrt{2}}( i\alpha_2 + \alpha_3 )
\;,
\eeq

\beq
\psi_{10} = \frac{1}{\sqrt{2}}( i\alpha_2 - \alpha_3 )
\;,
\eeq
\beq
\psi_{11} = \frac{1}{\sqrt{2}}( \alpha_0 - i \alpha_1 )
\;.
\eeq
Substituting these equations into the definition Eq.(\ref{eq:t-def}) of $t$
yields

\beq
t = \left| \sum_{j=0}^3 \alpha_j^2 \right |^2
\;.
\eeq
Suppose that $Q_j = |\alpha_j^2|$ and $\theta_j = {\rm phase}(\alpha_j^2)$
for $j\in Z_{0,3}$. Then $\sum_{j=0}^3 Q_j = 1$ and $t= | \sum_{j=0}^3 Q_j e^{i\theta_j}|^2$.
Thus $0 \leq t \leq 1$ and $t=1$ iff the $\theta_j$'s are all zero (i.e., the 
$\alpha^2_j$'s are all real). $t=0$ iff $E_F(\psi_{xy}) = 0$, and $t=1$ iff
$E_F(\psi_{xy})$ is maximum. This is why.
From Fig.\ref{fig:po-h-plot}, it is clear that $h(p_0(t)) = E_F(\psi_{xy})$ is a monotonically increasing
function of $t$ which goes from 0 to 1 as $t$ goes from 0 to 1.

So far we have discussed $E_F$ for pure states.
There are still many unsolved mysteries about $E_F$ for mixed states.
An example for which
definition Eq.(\ref{eq:ef-mix}) has been evaluated is 
when $\rho$ is diagonal in the Bell basis:

\beq
\rho = \sum_a w_a \ket{B_a}\bra{B_a}
\;,
\eeq
where the $w_a$'s are non-negative numbers that add up to one.
Ref.\cite{MonsterHootters} shows that for this $\rho$,

\beq
E_F(\rho) = \left\{
\begin{array}{ll}
0&{\rm if} \;\; W\leq \frac{1}{2}\\
h\left( \frac{1 + \sqrt{1 - 4 (W - \frac{1}{2})^2}}{2}\right)&{\rm otherwise}
\end{array}
\right.
\;,
\label{eq:entagf-bell-b}\eeq
where

\beq
W = \max_a (w_a)
\;.
\eeq

\BeginSection{Some Definitions}\label{sec:defs}

In this section, we will define our measure of entanglement. 
Future sections will explore the properties of our measure,
and how it compares with $E_F$.

Consider either a QB or CB net 
with $N$ nodes $(\rvx.)_\zn$.
Suppose that $\rvL_1, \rvL_2, \ldots, \rvL_n$
and $\rvE$ are non-empty disjoint node collections of the net.
For a CB net, we define the {\it H-tanglement} $HT$ for $n$
listeners (or receivers) $\rvL_1, \rvL_2, \ldots, \rvL_n$
and a speaker (or sender) $\rvE$ by

\beq
HT(\rvL_1 \tsep \rvL_2 \tsep \ldots \tsep \rvL_n | \rvE) =
\sum_{i=1}^n H(\rvL_i | \rvE) - H(\rvL_1, \rvL_2, \ldots, \rvL_n | \rvE)
\;.
\eeq
Analogously, for a QB net we define the {\it S-tanglement} $ST$ by

\beq
ST_\rho(\rvL_1 \tsep \rvL_2 \tsep  \ldots \tsep \rvL_n | \rvE) =
\sum_{i=1}^n S_\rho(\rvL_i | \rvE) - S_\rho(\rvL_1, \rvL_2, \ldots, \rvL_n | \rvE)
\;.
\eeq
Here $\rho$ is any density matrix obtained by reducing the meta density 
matrix of the net, but such that the nodes in
$\rvL_1, \rvL_2, \ldots, \rvL_n$
and  $\rvE$ haven't been reduced.
We will also use the term {\it max S-tanglement} to refer to $ST$ 
maximized over all local unitary
operations on the $\rvL_i$'s. 
If $ST_\rho \neq 0$ for a QB net but $HT=0$ for its parent CB net,
we will describe this situation by saying that there is {\it non-classical tanglement}.
$H(\rvL_1 : \rvL_2 : \ldots :\rvL_n)$ 
( or $S(\rvL_1 : \rvL_2 : \ldots :\rvL_n)$ ) will be called an 
{\it H (or S) mutual  information} for $n$ parts. $H(\rvL_1 : \rvL_2 : \ldots :\rvL_n |\rvE)$ 
( or $S(\rvL_1 : \rvL_2 : \ldots :\rvL_n | \rvE)$ ) will be called an 
{\it H (or S) conditional mutual information (c.m.i.)} for $n$ listeners.
When there are two listeners, tanglement equals a c.m.i..
As we shall see later, this is no longer the case for more than two listeners.

Recall from Ref.\cite{TucciQInfo} that
a node collection with more than one node is said to be {\it compound}.
Likewise, a listener or speaker with more than one node will be said to be compound.

Suppose that $\rvX$ and $\rvY$ are non-empty disjoint node 
collections of either a CB or a QB net.
For a CB net, we will say that 
$\rvX$ and $\rvY$ are {\it (probabilistically) independent} (also called 
separable or uncorrelated) if 

\beq
P(X, Y) = P(X) P(Y)
\;,
\eeq
for all possible $X$ and $Y$.
For a QB net, suppose $\rho_{\rvX, \rvY}$
is a density matrix acting on $\hil_{\rvX, \rvY}$
and obtained by reducing the meta density matrix of the net.
We will say that $\rvX$ and $\rvY$ are independent (or separable) if 

\beq
\rho_{\rvX, \rvY} = \rho_\rvX \;\;\rho_\rvY
\;.
\eeq

Suppose that $\rvX$, $\rvY$ and $\rvE$ are non-empty disjoint node 
collections of either a CB or a QB net.
For a CB net, we will say that 
$\rvX$ and $\rvY$ are {\it conditionally independent} (or conditionally separable) if 

\beq
P(X, Y) = \sum_E P(X|E) P(Y|E) P(E)
\;,
\eeq
for all possible $X$ and $Y$.
For a QB net, suppose $\rho_{\rvX, \rvY, \rvE}$
is a density matrix acting on $\hil_{\rvX, \rvY, \rvE}$
and obtained by reducing the meta density matrix of the net.
We will say that $\rvX$ and $\rvY$ are conditionally independent (or conditionally separable) if 

\beq
\rho_{\rvX, \rvY, \rvE} = \sum_E \rho_\rvX^{(E)} \rho_\rvY^{(E)} w_E \ket{E}\bra{E}
\;,
\eeq
where $\{ \ket{E} | \forall E \}$ is orthonormal basis corresponding to $\rvE$,
$w_E\geq 0$ for all $E$, $\sum_E w_E = 1$, 
$\rho_\rvX^{(E)}$ acts on $\hil_\rvX$,
and $\rho_\rvY^{(E)}$ acts on $\hil_\rvY$.

\BeginSection{$ST$ for 2 Single-node Listeners and a Pure State}\label{sec:st-2-pure}

In this section, we will discuss S-tanglement for 2 single-node listeners and a pure state.
We will show that it equals $E_F$ if we maximize it over all local unitary 
transformations on the two listeners. 

\begin{figure}[h]
	\begin{center}
	\epsfig{file=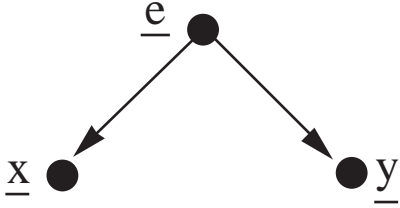}
	\caption{Net for 2 single-node listeners and a pure state.}
	\label{fig:2list-pure}
	\end{center}
\end{figure}

Consider the QB net of Fig.\ref{fig:2list-pure}, where
\BeginBNetTabular
	$\rve$ & $e = (e_1, e_2)$ & $\psi(e) = \sum_{x',y'} U_{e_1 x'} \psi^0(x',y') V^\dagger_{y' e_2}$ 
	& $\sum_{x,y} |\psi^0(x,y)|^2 = 1,$\\
	&&&$\sum_a U^*_{ax}U_{ax'} = \delta^x_{x'}$, \\
	&&& $\sum_b V^*_{by}V_{by'} = \delta^y_{y'}$\\
\hline
	$\rvx$ & $x\in S_\rvx$ & $\delta(x, e_1)$ & \\
\hline
	$\rvy$ & $y\in S_\rvy$ & $\delta(y, e_2)$ & \\
\EndBNetTabular
We will sometimes write $\psi_{xy}$ instead of $\psi(x,y)$. Without loss of generality,
we will assume that $N_\rvx$ (the size of set $S_\rvx$) is less than
or equal to $N_\rvy$.

The meta density matrix $\mu$ of this net is

\beq
\mu = \rhometa
\;,
\eeq
where

\beq
\ketmeta = \sum_{ri} \psi(x, y) \ket{ \rve = (x, y), x, y}
\;.
\eeq
Define $\rho$ by

\beq
\rho = \trace{\rve} (\mu) = 
\sum_{ri} \psi(x, y) \psi^*(x, y) \ket{ x, y} \bra{ x, y}
\;.
\eeq
One has that 

\beq
S_\mu(\rvx : \rvy | \rve)
= S_\mu(\rvx ,\rve)
+ S_\mu(\rvy ,\rve)
- S_\mu(\rvx , \rvy, \rve)
- S_\mu(\rve)
\;.
\label{eq:sxy-e}\eeq
But $\mu$ is a pure state acting on $\hil_{\rvx, \rvy, \rve}$, so

\begin{subequations}
\beq
S_\mu(\rvx ,\rve) = S_\mu(\rvy)
\;,
\eeq

\beq
S_\mu(\rvy ,\rve) = S_\mu(\rvx)
\;,
\eeq

\beq
S_\mu(\rvx , \rvy, \rve) = 0
\;,
\eeq

\beq
S_\mu(\rve) = S_\mu(\rvx , \rvy)
\;.
\eeq
\label{eq:pure-simplify}\end{subequations}
Substituting Eqs.(\ref{eq:pure-simplify}) into Eq.(\ref{eq:sxy-e})
yields

\beq
S_\mu(\rvx:\rvy | \rve) = S_\mu(\rvx : \rvy) = S_\rho(\rvx : \rvy)
\;.
\label{eq:intermediate}\eeq
Note that $\rho$ is diagonal in the $\ket{x,y}$ basis
so Eq.(\ref{eq:intermediate}) can  be simplified further.
Let 

\beq
P(x,y) = |\psi(x, y)|^2
\;.
\eeq
With this $P(x,y)$,
one can calculate $H(\rvx : \rvy)$. Eq.(\ref{eq:intermediate})
reduces to

\beq
S_\mu(\rvx:\rvy | \rve) = H(\rvx : \rvy)
\;.
\label{eq:s-equals-hxy}\eeq

Henceforth, we will often abbreviate $P(x, y)$ by $P_{xy}$, 
$P(x) = \sum_y P(x,y)$ by $P_{x-}$, and  
$P(y) = \sum_x P(x,y)$ by $P_{-y}$.

When $S_\rvx =S_\rvy = Bool$, the unitary matrices $U$ and $V$ mentioned in the above table
determine what spin direction is measured at the nodes $\rvx$ and $\rvy$.
The above table and the following one
\BeginBNetTabular
	$\rve$ & $e = (e_1, e_2)$ & $\psi^0(e)$ & \\
\hline
	$\rvx$ & $x\in S_\rvx$ & $U_{e_1, x}$ & \\
\hline
	$\rvy$ & $y\in S_\rvy$ & $V^\dagger_{y, e_2}$ & \\
\EndBNetTabular
do not yield the same $S_\mu(\rvx:\rvy | \rve)$.
In the first table, node $\rve$ upon which we
condition has knowledge of $U$ and $V$, whereas in the second
it doesn't. We will call the $U$ and $V$ in the first (ditto, second) table
{\it a priori (ditto, a posteriori) local unitary transformations} on $\rvx$ and $\rvy$.
In this section, we are interested in the case of the first table,
where $U$ and $V$ refer to a priori transformations.

Suppose $\psi$ (ditto, $\psi^0$) is the rectangular matrix 
with entries $\psi_{xy}$ (ditto, $\psi^0_{xy}$). Then 

\beq
\psi = U \psi^0 V^\dagger
\;.
\eeq
Let us consider the special case that $U$ and $V$ make $\psi$ diagonal. 
Such a $U$ and $V$  always exist 
by the
Singular Value Decomposition Theorem. Suppose that 

\beq
\psi\psi^\dagger = diag(p_0, p_1 ,\ldots, p_{N_\rvx -1})
\;.
\eeq
The $p_x$'s must be non-negative numbers that add up to one. 
Then

\beq
H(\rvx : \rvy) = \sum_{x,y} P_{xy} \log_2 \frac{P_{xy}}{P_{x-} P_{-y}}
=
\sum_x p_x \log_2 \frac{1}{p_x} = E_F(\psi^0_{xy}) = E_F(\psi_{xy})
\;.
\eeq
Combining the last equation and Eq.(\ref{eq:s-equals-hxy}) yields

\beq
S_\mu(\rvx : \rvy | \rve) = E_F(\psi_{xy})
\;
\label{eq:s-equals-ef}\eeq
for the special case that $U$ and $V$ make $\psi$ diagonal.

In Appendices \ref{app:ef-ineq} and \ref{app:mut-info-bound}, 
we show the following inequalities:

\beq
E_F(|\psi_{xy}|) \leq E_F(\psi_{xy})
\;,
\eeq

\beq
H(\rvx : \rvy) \leq E_F(|\psi_{xy}|)
\;.
\eeq
Combining these inequalities and Eq.(\ref{eq:s-equals-hxy})
yields

\beq
S_\mu(\rvx : \rvy | \rve) \leq E_F(\psi_{xy})
\;.
\label{eq:s-leq-ef}\eeq

From the argument leading up to Eq.(\ref{eq:s-equals-ef}), we see that there exists a pair of 
unitary matrices $U$ and $V$ so that the S-tanglement $ST$ 
equals the corresponding entanglement of formation $E_F$.
From the argument leading up to Eq.(\ref{eq:s-leq-ef}), we see that for any $U$ and $V$,
$ST$ is less than or equal to the corresponding $E_F$.
Therefore, if $ST$ is maximized over all a priori
local unitary transformations $U$ and $V$ on its two listeners,
then it equals $E_F$.

\BeginSection{$ST$ for 2 Single-Node Listeners and a Mixed State}\label{sec:st-2-mix}

In this section, we will discuss S-tanglement for 2 single-node listeners and a mixed state.
We will show that it vanishes for a conditionally separable state. We will also
calculate $ST$ for any $\rho$ which is diagonal in the Bell basis. 

Suppose $\rvq_1, \rvq_2, \rve$ are nodes of a QB net. Suppose

\beq
\rho = \sum_a w_a \rho_a^{(1)} \rho_a^{(2)}
\;,
\label{eq:proj-cond-separ}\eeq
where $w_a\geq 0$ for all $a$ and $\sum_a w_a =1$, and where for all $a$
and for $\lam=1,2$, 
$\rho_a^{(\lam)}$ is a density matrix acting on $\hil_{\rvq_\lam}$.
For such a $\rho$, $E_F(\rho)= 0$ \cite{MonsterHootters}.
To calculate $S_\rho(\rvq_1 : \rvq_2 | \rva)$, we 
need a $\rho$ that acts on a space $\hil_{\rvq_1, \rvq_2, \rva}$
or larger, so the $\rho$ in Eq.(\ref{eq:proj-cond-separ}) will not do.
Suppose we consider instead the following $\rho$:

\beq
\rho = \sum_a w_a \ket{a}\bra{a}\rho_a^{(1)} \rho_a^{(2)}
\;,
\label{eq:cond-separ}\eeq
where $\{ \ket{a} | \forall a \}$ is the orthonormal basis for node $\rva$.
For this $\rho$, one has

\beq
S_\rho(\rvq_1 : \rvq_2 | \rva) = 
 S_\rho(\rvq_1, \rva) 
+S_\rho(\rvq_2, \rva) 
-S_\rho(\rvq_1, \rvq_2, \rva) 
-S_\rho(\rva) 
\;,
\eeq
where

\beq
S_\rho(\rvq_\lam, \rva) = H(\vec{w}) + \sum_a w_a S(\rho_a^{(\lam)})\;\;\;\;\;\;{\rm for}\;\lam=1,2
\;,
\eeq

\beq
S_\rho(\rvq_1, \rvq_2, \rva) = H(\vec{w}) + \sum_a w_a \{ S(\rho_a^{(1)}) + S(\rho_a^{(2)})\}
\;,
\eeq

\beq
S_\rho(\rva) = H(\vec{w}) 
\;,
\eeq
so

\beq
S_\rho(\rvq_1 : \rvq_2 | \rva) =0
\;.
\eeq

\begin{figure}[h]
	\begin{center}
	\epsfig{file=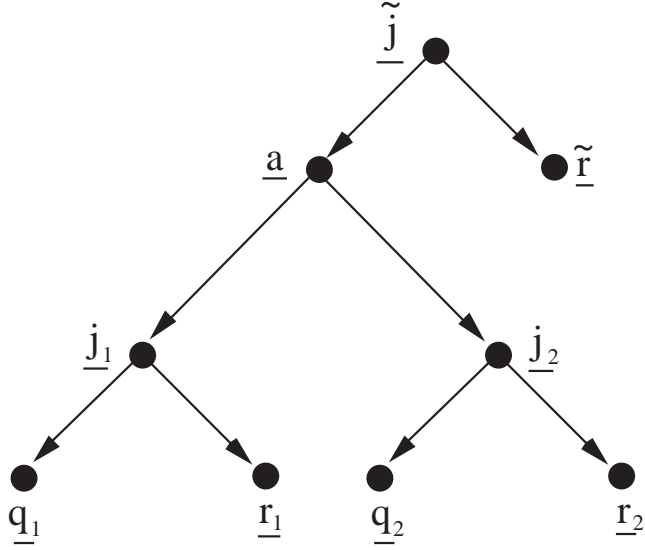, height=3in}
	\caption{Net that implements a 2 choice conditionally separable density matrix.}
	\label{fig:2list-cond-separ}
	\end{center}
\end{figure}

Note that the $\rho$ defined by Eq.(\ref{eq:cond-separ}) can be implemented
by the QB net of Fig.\ref{fig:2list-cond-separ}, where
\BeginBNetTabular
	$\markrvj$ & $\markj = (\markj^1, \markj^2)$ & $\sqrt{w_{\markj^1}}\delta(\markj^1, \markj^2)$ & $\sum_{\markj^1} w_{\markj^1} = 1$\\
\hline
	$\rva$ & $a$ & $\delta(a, \markj^1)$ & \\
\hline
	$\markrvr$ & $\markr$ & $\delta(\markr, \markj^2)$ & \\
\hline
	$\rvj_\lam$ for $\lam \in Z_{1,2}$ & $j_\lam = (j^1_\lam, j^2_\lam)$ & $\alpha_\lam(j_\lam | a)$ &
	$\sum_{j_\lam} |\alpha_\lam(j_\lam | a)|^2 = 1$\\
\hline
	$\rvq_\lam$ for $\lam \in Z_{1,2}$ & $q_\lam$ & $\delta(q_\lam, j^1_\lam)$ & \\
\hline
	$\rvr_\lam$ for $\lam \in Z_{1,2}$ & $r_\lam$ & $\delta(r_\lam, j^2_\lam)$ & \\
\EndBNetTabular
The meta density matrix $\mu$ of this net is 

\beq
\mu = \rhometa
\;,
\eeq
where

\beq
\ketmeta = \sum_{ri} \sqrt{w_a} 
\left[
\prod_{\lam=1}^2 \alpha_\lam(q_\lam, r_\lam | a) \ket{\rvj_\lam=(q_\lam, r_\lam), q_\lam, r_\lam}
\right]
\ket{\markrvj = (a, a), a, \markrvr = a}
\;.
\eeq
Define $\rho$ by

\beq
\rho =
\esum{\markrvj, \rvj_1, \rvj_2}
\trace{\markrvr, \rvr_1, \rvr_2} (\mu)
\;.
\eeq
Then

\beq
\rho =
\sum_a w_a \ket{a}\bra{a}\rho_a^{(1)} \rho_a^{(2)}
\;,
\eeq
where 

\beq
\rho_a^{(\lam)} = 
\sum_{all/a, \lam} 
\alpha_\lam(q_\lam, r_\lam | a)
\alpha^*_\lam(q'_\lam, r_\lam | a)
\ket{q_\lam}\bra{q'_\lam}
\;
\eeq
for all $a$ and for $\lam=1,2$.

\begin{figure}[h]
	\begin{center}
	\epsfig{file=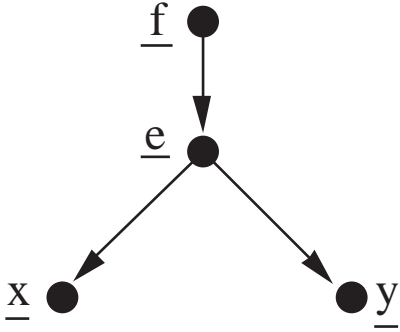}
	\caption{Net for 2 single-node listeners and a mixed state.}
	\label{fig:2list-mixd}
	\end{center}
\end{figure}

Next consider the QB net of Fig.\ref{fig:2list-mixd}, where
\BeginBNetTabular
	$\rvf$ & $f$ & $\sqrt{w_f}$ & $\sum_f w_f = 1$\\
\hline
	$\rve$ & $e=(e_1, e_2)$ & $\av{e| \psi_f} = \psi_f(e)$ & $\sum_{e} |\psi_f (e)|^2= 1$\\
\hline
	$\rvx$ & $x$ & $\delta(x, e_1)$ & \\
\hline
	$\rvy$ & $y$ & $\delta(y, e_2)$ & \\
\EndBNetTabular
The meta density matrix $\mu$ of this net is 

\beq
\mu = \rhometa
\;,
\eeq
where

\beq
\ketmeta = \sum_{ri} \sqrt{w_f} \psi_f(x,y)
\ket{f, \rve=(x,y), x, y}
\;.
\eeq
Define $\sigma$ by

\beq
\sigma = \trace{f} (\mu) =
\sum_{ri} w_f \psi_f(x,y) \psi^*_f(x',y')
\ket{\rve=(x,y), x, y}\bra{\rve=(x',y'), x', y'}
\;.
\eeq
We wish to calculate $S_\sigma(\rvx : \rvy | \rve)$.
Let 

\beq
P(x, y) = \sum_f w_f |\psi_f (x,y)|^2
\;.
\eeq
We can define a density matrix $\rho(y)$ for each $y\in S_\rvy$ by

\beq
\rho(y) =
\frac{ \sum_{f, x, x'} w_f \psi_f(x, y) \psi_f(x', y) \ket{x}\bra{x'}}{P(y)}
\;.
\eeq
In an analogous manner, we can define a density matrix $\rho(x)$ for each $x\in S_\rvx$. It is also
convenient to define $\rho$ by

\beq
\rho = \esum{\rve}\trace{\rvf}(\mu) = \sum_f w_f \ket{\psi_f}\bra{\psi_f}
\;.
\eeq
One has that 

\beq
S_\sigma(\rvx : \rvy | \rve) =
  S_\sigma(\rvx , \rve)
+ S_\sigma(\rvy, \rve)
- S_\sigma(\rvx, \rvy, \rve)
- S_\sigma(\rve)
\;.
\eeq
Using the observations of Appendix \ref{app:rep-indices},
one gets

\beq
S_\sigma(\rvx, \rve) =
S\left[\sum_y P(y) \ket{y}\bra{y} \rho(y)\right] =
H(\rvy) + \sum_y P(y) S[\rho(y)]
\;.
\eeq
Likewise,

\beq
S_\sigma(\rvy, \rve) =
H(\rvx) + \sum_x P(x) S[\rho(x)]
\;.
\eeq
Furthermore,

\beq
S_\sigma(\rvx, \rvy, \rve) = S(\rho)
\;,
\eeq
and

\beq
S_\sigma(\rve) = H(\rvx, \rvy)
\;.
\eeq
Therefore,

\beq
S_\sigma(\rvx : \rvy | \rve) =
H(\rvx : \rvy)
+ \sum_x P(x) S[\rho  (x)]
+ \sum_y P(y) S[\rho  (y)]
- S(\rho)
\;.
\label{eq:gen-st-formula}\eeq

Note that if $w_f = \delta(f,0)$, then $\rho(x)$, $\rho(y)$
and $\rho$ are all pure states so the right-hand side of the last equation 
reduces to $H(\rvx : \rvy)$. This is what the previous section
on pure states would lead us to expect. 

Now consider the case that $S_\rvx= S_\rvy= Bool$.
Let $w_{x-} = \sum_{y=0}^1 w_{xy}$, and
$w_{-y} = \sum_{x=0}^1 w_{xy}$.
If we specialize Eq.(\ref{eq:gen-st-formula})
by assuming that the states $\ket{\psi_f}$ are the Bell Basis states (defined by 
Eq.(\ref{eq:bell-basis}) ), then we obtain

\beq
S_\sigma(\rvx : \rvy | \rve) = h(w_{0-}) + 1 - H(\vec{w})
\;.
\label{eq:st-bell-mix}\eeq
The last equation gives $ST$ for a 
Bell diagonal mixture.  
$E_F(\rho)$ for this same state was given 
in Eq.(\ref{eq:entagf-bell-b}). 
I'm not sure yet how these two results are connected.
Also, note that Eq.(\ref{eq:st-bell-mix}) is not yet 
maximized over all a priori local unitary transformations, and
one should perform this maximization before comparing it with $E_F(\rho)$, if
one is to follow the same rules that were used in the 
pure state case.

\BeginSection{Properties of Tanglement and C.M.I.}\label{sec:props}
In this section we will discuss various properties
satisfied by tanglements and c.m.i.'s.

The following notation will be used henceforth.

Often, after stating something about the classical 
entropy $H$ or the classical tanglement $HT$, we will
append to the end of the statement the symbol \HS
to indicate that the statement is also valid
if one replaces $H$ by $S$ everywhere.
Likewise, the symbol \SH will indicate that the previous statement
is also valid if we replace $S$ by $H$ everywhere.

 For any set $S$,
its power set $Pow(S)$ is the set of all subsets of $S$,
including the null set. For example,
$Pow(\{1,2\}) = \{ \emptyset, \{1\}, \{2\}, \{1,2\}\}$
If $S$ has $|S|$ elements, then
$Pow(S)$ has $2^{|S|}$ elements. For this reason $Pow(S)$  is
often denoted by $2^S$. We will also use $Pow(S)_j$ for any
$j\in Z_{0,|S|}$ to denote the set of all subsets of $S$
which contain $j$ elements. 
For example,
$Pow(Z_{1,3})_2 = \{ \{1,2\}, \{1,3\}, \{2,3\}\}$
Clearly, $Pow(S) = \cup_{j=0}^{|S|} Pow(S)_j$.

%For any set $S = \{ a_1, a_2, \ldots, a_n\}$, let
%$S^: = a_1 : a_2 : \ldots : a_n$, and 
%$S^\tsep = a_1 \tsep a_2 \tsep \ldots \tsep a_n$.
%In other words, a set with superscript $\tsep$ (ditto, $:$)
%will denote the elements of the set separated by $\tsep$ (ditto, $:$)
%instead of commas.

For any set $S = \{ a_1, a_2, \ldots, a_n\}$, let
%$(\tsep_{a\in S} a)= (\tsep_{j=1}^n a_j) = a_1 \tsep a_2 \tsep \ldots \tsep a_n$, and 
$(:_{a\in S} a)= (:_{j=1}^n a_j) = a_1 : a_2 : \ldots : a_n$.

Suppose $\rvE, \rvX_1, \rvX_2, \ldots, \rvX_n$ with $n\geq 2$
are non-empty disjoint node collections of a Bayesian net,
and $\Gamma_\alpha$ for $\alpha\in Z_{1,m}$ are
non-empty disjoint subsets of $Z_{1,n}$. 
We will sometimes use the following $\tau, \mu$ shorthand for tanglement and 
c.m.i.:

\beq
\tau(\Gamma_1 \tsep \Gamma_2 \tsep \ldots \tsep \Gamma_m) =
HT[ (\rvX.)_{\Gamma_1} \tsep (\rvX.)_{\Gamma_2} \tsep \ldots \tsep (\rvX.)_{\Gamma_m} | \rvE ] 
\;,
\mathHS\eeq

\beq
\mu(\Gamma_1 : \Gamma_2 : \ldots : \Gamma_m) =
H[  (\rvX.)_{\Gamma_1} : (\rvX.)_{\Gamma_2} : \ldots : (\rvX.)_{\Gamma_m} | \rvE ] 
\;.
\mathHS\eeq
For example,

\beq
\tau(1 \tsep 2 \tsep (3,4)) = HT(\rvX_1 \tsep \rvX_2 \tsep (\rvX_3, \rvX_4)| \rvE)
\;,
\mathHS\eeq

\beq
\mu(1 : 2 : (3,4)) = H(\rvX_1: \rvX_2 : (\rvX_3, \rvX_4) | \rvE)
\;.
\mathHS\eeq
Sometimes, we will put the argument of $\tau$ or $\mu$ in a subscript (e.g., $\tau_{1\tsep 2}$), while other times
we will put it in parentheses (e.g., $\tau(1\tsep 2)$).

In discussing the following properties, we will use $\rvE, \rvX_1, \rvX_2, \ldots, \rvX_n$ with $n\geq 2$
to denote non-empty disjoint node collections of a Bayesian net.

\MyCases{(1) Symmetry}

H \HS tanglement and c.m.i. are symmetric under permutations of their listeners.

\MyCases{(2) Sign of tanglement}

One has that

\beq
H(\rvX_1 : \rvX_2 | \rvE)
= H(\rvX_1 | \rvE) + H(\rvX_2 | \rvE) - H[(\rvX_1, \rvX_2) | \rvE]
= H(\rvX_1 | \rvE) - H(\rvX_1 | \rvE, \rvX_2)
\geq 0
\;.
\mathHS\eeq
where the inequality follows by strong subadditivity.

Tanglement is non-negative for any number of listeners, not just two.
Indeed, an $n$-listener tanglement can always be expressed as a sum of
2-listener tanglements.
For example, for 4 listeners, one has

\beq
\tau(1\tsep 2 \tsep 3 \tsep 4) =
\tau( 1 \tsep 2) 
+ \tau ((1,2)\tsep3)
+\tau( (1,2,3) \tsep 4) \geq 0
\;.
\mathHS\eeq

\MyCases{(3) Decomposition of c.m.i.}

In discussing tanglements, c.m.i.'s
often arise. Next we will show how to express a c.m.i.
as a sum of $\pm$  non-mutual informations.

For 2 listeners

\beq
H(\rvX_1 : \rvX_2 | \rvE) = H(\rvX_1 | \rvE) + H(\rvX_2 | \rvE) - H(\rvX_1, \rvX_2 | \rvE)
\;,
\mathHS\eeq

\beq
H(\rvX_1 : \rvX_2 | \rvE) = H(\rvX_1 , \rvE)  + H(\rvX_2 , \rvE) - H(\rvX_1, \rvX_2 , \rvE) - H(\rvE)
\;.
\mathHS\eeq
For 3 listeners,

\beq
H(\rvX_1 : \rvX_2 : \rvX_3 | \rvE) =
\left\{ 
\begin{array}{l}
H(\rvX_1 | \rvE) + H(\rvX_2 | \rvE) + H(\rvX_3 | \rvE)\\
- H(\rvX_1, \rvX_2 | \rvE) - H(\rvX_1, \rvX_3 | \rvE) - H(\rvX_2, \rvX_3 | \rvE)\\
+ H(\rvX_1, \rvX_2, \rvX_3 | \rvE)
\end{array}
\right.
\;,
\label{eq:mut-info-3l}
\mathHS\eeq

\beq
H(\rvX_1 : \rvX_2 : \rvX_3 | \rvE) =
\left\{ 
\begin{array}{l}
H(\rvX_1 , \rvE) + H(\rvX_2 , \rvE) + H(\rvX_3 , \rvE)\\
- H(\rvX_1, \rvX_2 , \rvE) - H(\rvX_1, \rvX_3 , \rvE) - H(\rvX_2, \rvX_3 , \rvE)\\
+ H(\rvX_1, \rvX_2, \rvX_3 , \rvE)\\
- H(\rvE)
\end{array}
\right.
\;.
\mathHS\eeq
For 4 listeners,

\beq
H(\rvX_1 : \rvX_2 : \rvX_3 : \rvX_4 | \rvE) =
\left\{ 
\begin{array}{l}
\sum_{\alpha=1}^4 
H(\rvX_\alpha | \rvE)\\
- \sum_{1\leq \alpha < \beta \leq 4} H(\rvX_\alpha, \rvX_\beta | \rvE)\\
+ \sum_{1\leq \alpha < \beta < \gamma \leq 4} H(\rvX_\alpha, \rvX_\beta, \rvX_\gamma | \rvE)\\
-H(\rvX_1, \rvX_2, \rvX_3, \rvX_4| \rvE)
\end{array}
\right.
\;,
\mathHS\eeq

\beq
H(\rvX_1 : \rvX_2 : \rvX_3 : \rvX_4 | \rvE) =
\left\{ 
\begin{array}{l}
\sum_{\alpha=1}^4 
H(\rvX_\alpha , \rvE)\\
- \sum_{1\leq \alpha < \beta \leq 4} H(\rvX_\alpha, \rvX_\beta , \rvE)\\
+ \sum_{1\leq \alpha < \beta < \gamma \leq 4} H(\rvX_\alpha, \rvX_\beta, \rvX_\gamma , \rvE)\\
-H(\rvX_1, \rvX_2, \rvX_3, \rvX_4 , \rvE)\\
-H(\rvE)
\end{array}
\right.
\;.
\mathHS\eeq
One can show by induction that for $n\geq 2$ listeners,

\beq
H(:_{\lam=1}^n \rvX_\lam |\rvE) = \sum_{\lam=1}^{n} (-1)^{\lam+1} \sum_{\Gamma\in Pow(Z_{1,n})_\lam} H[ (\rvX.)_\Gamma | \rvE]
\;,
\mathHS\eeq

\beq
H(:_{\lam=1}^n \rvX_\lam |\rvE) = 
\left\{ 
\begin{array}{l}
\sum_{\lam=1}^{n} (-1)^{\lam+1} \sum_{\Gamma\in Pow(Z_{1,n})_\lam} H[ (\rvX.)_\Gamma, \rvE] \\
-H(\rvE)
\end{array}
\right.
\;.
\mathHS\eeq

For the quantum case, a simple consequence of the above decomposition of c.m.i. is as follows.
For 2 listeners, 

\beq
S_\rho(\rvX_1, \rvX_2, \rvE) = 0\;\;{\rm implies}\;\;
S_\rho(\rvX_1: \rvX_2 | \rvE) = S_\rho(\rvX_1, \rvX_2)
\;.
\eeq
For 3 listeners, 

\beq
S_\rho(\rvX_1, \rvX_2, \rvX_3, \rvE) = 0\;\;{\rm implies}\;\;
S_\rho(\rvX_1: \rvX_2 : \rvX_3 | \rvE) = -S_\rho(\rvX_1 : \rvX_2 : \rvX_3)
\;.
\eeq
One can show that for $n\geq 2$ listeners,

\beq
S_\rho(\rvX_1, \rvX_2, \ldots, \rvX_n, \rvE) = 0\;\;{\rm implies}\;\;
S_\rho(\rvX_1: \rvX_2 : \ldots : \rvX_n | \rvE) = (-1)^n S_\rho(\rvX_1 : \rvX_2 : \ldots : \rvX_n)
\;.
\eeq

\MyCases{(4) Sign of c.m.i.}

The c.m.i. $H(\rvX_1: \rvX_2: \ldots : \rvX_n | \rvE)$ \HS
is non-negative for $n=2$,
because in that case it equals the tanglement $HT(\rvX_1 \tsep \rvX_2 | \rvE)$ \HS.
However, for more than 2 listeners, the c.m.i. 
may be positive or negative, as the following example shows.\cite{McGill}
A 3 listener c.m.i. will be positive if one of the 3 listeners
drops out so that there are effectively 2 listeners.  
Let us construct an example of a 3 listener c.m.i. that is negative.
Assume the listeners are independent of 
the speaker $\rvE$ so that we can omit the conditioning on $\rvE$.
Eq.(\ref{eq:mut-info-3l}) can be rewritten as

\beq
H(\rvX_1: \rvX_2: \rvX_3) =
Pos + Neg
\;,
\eeq
where

\beq
Pos = H(\rvX_1) - H(\rvX_1 | \rvX_2) = H(\rvX_1 : \rvX_2)
\;,
\eeq
and

\beq
Neg = -\{ H(\rvX_1 | \rvX_3) - H(\rvX_1 | \rvX_2, \rvX_3) \} = - H[(\rvX_1 : \rvX_2) | \rvX_3]
\;.
\eeq
As their names suggest, $Pos$ and $Neg$ are positive and negative, respectively.
The idea is to make $\rvX_1$ and $\rvX_2$ independent
so that $Pos$ vanishes. The following probability distribution fits that bill:

\beq
P(X_1, X_2, X_3) =
\frac{1}{4} [ 
\delta^{X_1}_{0} \delta^{X_2}_{\bar{X_3}}
+
\delta^{X_1}_{1} \delta^{X_2}_{X_3}
]
\;,
\eeq
where $X_1, X_2, X_3 \in Bool$, $\bar{0} = 1$ and $\bar{1} = 0$. This distribution 
gives $Pos = 0$ and $Neg= -1$.

\MyCases{(5) Duality between tanglement and c.m.i.}

We wish to express tanglements in terms of c.m.i.'s
and vice versa. For 2 listeners, one finds

\beq
\tau_{1 \tsep 2} = \mu_{1:2}
\;,
\mathHS\eeq
For 3 listeners, one finds

\beq
\tau_{1 \tsep 2 \tsep 3} = \mu_{1:2} + \mu_{1:3} + \mu_{2:3} - \mu_{1:2:3}
\;,
\mathHS\eeq

\beq
\mu_{1:2:3} = \tau_{1 \tsep 2} + \tau_{1 \tsep 3} + \tau_{2 \tsep 3} - \tau_{1 \tsep 2 \tsep 3}
\;.
\mathHS\eeq
For 4 listeners, one finds

\beq
\tau(1 \tsep 2 \tsep 3 \tsep 4) = 
 \sum_{\Gamma\in Pow(Z_{1,4})_2}\mu(:_{j\in \Gamma}j) 
-\sum_{\Gamma\in Pow(Z_{1,4})_3}\mu(:_{j\in \Gamma}j) 
+ \mu(1:2:3:4)
\;,
\mathHS\eeq

\beq
\mu(1:2:3:4) = 
 \sum_{\Gamma\in Pow(Z_{1,4})_2}\tau(\tsep_{j\in \Gamma}j)  
-\sum_{\Gamma\in Pow(Z_{1,4})_3}\tau(\tsep_{j\in \Gamma}j) 
+ \tau(1 \tsep 2 \tsep 3 \tsep 4)
\;.
\mathHS\eeq
One can show by induction that for $n\geq 2$ listeners

\beq
\tau(1\tsep 2\tsep \ldots \tsep n) =
\sum_{\lam=2}^n  (-1)^\lam \sum_{\Gamma\in Pow(Z_{1,n})_\lam} \mu(:_{j\in \Gamma}j) 
\;,
\label{eq:tau-mu}
\mathHS\eeq

\beq
\mu(1: 2: \ldots : n) =
\sum_{\lam=2}^n  (-1)^\lam \sum_{\Gamma\in Pow(Z_{1,n})_\lam} \tau(\tsep_{j\in \Gamma}j) 
\;.
\label{eq:mu-tau}
\mathHS\eeq
An interesting aspect of Eqs.(\ref{eq:tau-mu}) and Eqs.(\ref{eq:mu-tau}) is that
they transform into each other when one exchanges 
%the symbols $\tau$, semicolon with  $\mu$, colon. 
the symbols $\tau$ and  $\mu$. 
Therefore, we will call such equations {\it duality equations}, 
and say that they describe a 
{\it duality} between 
tanglement and c.m.i..

\MyCases{(6) Merging two listeners}

It is easy to check that for $n\geq 2$,

\beq
\begin{array}{l}
HT(\rvX_1 \tsep \rvX_2 \tsep \ldots \tsep \rvX_n \tsep \rvX_{n+1} | \rvE)
-  HT(\rvX_1 \tsep \rvX_2 \tsep \ldots \tsep \rvX_{n-1} \tsep (\rvX_n , \rvX_{n+1})| \rvE) = \\
\;\;=HT(\rvX_n \tsep \rvX_{n+1} | \rvE) \geq 0
\end{array}
\;.
\mathHS\eeq
In $\tau$ notation,
\beq
\tau[1 \tsep 2 \tsep \ldots \tsep n-1 \tsep (n, n+1)] \leq 
\tau[1 \tsep 2 \tsep \ldots \tsep n \tsep n+1]
\;.
\mathHS\eeq
For example,

\beq
\tau_{1\tsep 2,3} \leq \tau_{1\tsep 2\tsep 3}
\;.
\mathHS\eeq
Thus, ``merging" two listeners decreases tanglement.
Since tanglement is non-negative, if the right-hand side of this inequality is zero, so is the left-hand side.

\MyCases{(7) Pruning or removing a listener}

It is easy to check that for $n\geq 2$,

\beq
\begin{array}{l}
HT(\rvX_1 \tsep \rvX_2 \tsep \ldots \tsep \rvX_{n-1} \tsep (\rvX_n, \rvX_{n+1}) | \rvE)
-  HT(\rvX_1 \tsep \rvX_2 \tsep \ldots \tsep \rvX_{n}| \rvE) = \\
\;\;=HT[(\rvX_1 , \rvX_2 , \ldots , \rvX_{n-1}) : \rvX_{n+1} | \rvX_n, \rvE)\geq 0
\end{array}
\;.
\mathHS\eeq
In $\tau$ notation,

\beq
\tau[1\tsep 2\tsep \ldots \tsep n-1 \tsep n] \leq  \tau[1\tsep 2\tsep \ldots \tsep n-1 \tsep (n,n+1)]
\;.
\mathHS\eeq
For example,

\beq
\tau_{1\tsep 2} \leq  \tau_{1\tsep (2,3)}
\;.
\mathHS\eeq
Thus, ``pruning" a listener (i.e., removing some but not all of its nodes) decreases tanglement.
Since tanglement is non-negative, if the right-hand side of this inequality is zero, so is the left-hand side.

And what happens if we remove all the nodes of a listener? It is easy to check that
for $n\geq 2$,

\beq
\begin{array}{l}
HT(\rvX_1 \tsep \rvX_2 \tsep \ldots \tsep \rvX_n \tsep \rvX_{n+1}) | \rvE)
-  HT(\rvX_1 \tsep \rvX_2 \tsep \ldots \tsep \rvX_{n}| \rvE) = \\
\;\;=HT[(\rvX_1 , \rvX_2 , \ldots , \rvX_n) : \rvX_{n+1} | \rvE) \geq 0
\end{array}
\;.
\label{eq:remove-list}
\mathHS\eeq
In the $\tau$ notation, 

\beq
\tau(1\tsep 2\tsep \ldots \tsep n) 
\leq
\tau(1\tsep 2\tsep \ldots \tsep n \tsep n+1)
\;.
\mathHS\eeq
For example,

\beq
\tau_{1\tsep 2} \leq  \tau_{1\tsep 2 \tsep 3}
\;.
\mathHS\eeq
Thus, completely ``removing" a listener also decreases tanglement.
Since tanglement is non-negative, if the right-hand side of this inequality is zero, so is the left-hand side.

Note that if $\tau_{1 \tsep 2 \tsep \ldots \tsep n}=0$ for some $n$, then  $\mu_{1: 2: \ldots : n}=0$.
Indeed, by the duality equations, $\mu_{1: 2: \ldots : n}$ can be expressed as a sum of  $\pm$
$\tau$'s obtained from $\tau_{1\tsep 2\tsep \ldots \tsep n}$ by removing some of
its listeners. But all such $\tau$ must be zero because $\tau_{1\tsep 2\tsep \ldots \tsep n}=0$
and removing listeners decreases tanglement.

\MyCases{(8) Decomposing compound listeners of tanglement and c.m.i. } 

It is easy to check that
\beq
\tau_{1 \tsep 2,3} = \tau_{1 \tsep 2 \tsep 3} - \tau_{2 \tsep 3}
\;,
\mathHS\eeq

\beq
\tau_{1,2 \tsep 3,4}
=
\tau_{1 \tsep 2 \tsep 3 \tsep 4}
-\tau_{1 \tsep 2}
-\tau_{3 \tsep 4}
\;,
\mathHS\eeq

\beq
\tau_{1 \tsep 2,3 \tsep 4,5,6}
=
\tau_{1 \tsep 2 \tsep 3 \tsep 4 \tsep 5}
-\tau_{2 \tsep 3}
-\tau_{4 \tsep 5 \tsep 6}
\;.
\mathHS\eeq
Note that 
compound listeners
in the left-hand side 
are ``split" in the right-hand side.
More generally, suppose that
$\rvE, \rvX_1, \rvX_2, \ldots, \rvX_n$ for some $n\geq 2$
are non-empty disjoint node collections of a Bayesian net,
and $\Gamma_\alpha$ for $\alpha\in Z_{1,m}$ are
non-empty disjoint subsets of $Z_{1,n}$. Then 

\beq
HT[ \tsep_{\alpha=1}^m (\rvX.)_{\Gamma_\alpha} | \rvE] =
HT[ \tsep_{j \in \Gamma_1 \cup \Gamma_2 \ldots \Gamma_m} (\rvX.)_j | \rvE] 
-\sum_{\alpha=1}^{m} HT[ \tsep_{j \in \Gamma_\alpha}\rvX_j | \rvE]
\;,
\mathHS\eeq
where we define $HT[ \tsep_{j \in \Gamma_\alpha}\rvX_j | \rvE]=0$ if $\Gamma_\alpha$ has only one element.
In $\tau$ notation,

\beq
\tau(\Gamma_1 \tsep \Gamma_2 \tsep \ldots \tsep \Gamma_m)
=\tau(\tsep_{j \in \Gamma_1 \cup \Gamma_2 \ldots \Gamma_m} j ) 
-\sum_{\alpha =1}^m \tau(\tsep_{j \in \Gamma_\alpha} j )
\;,
\label{eq:split-comp-list}
\mathHS\eeq
where we define $\tau(\tsep_{j \in \Gamma_\alpha} j )=0$ if $\Gamma_\alpha$ has only one element.
Thus, any tanglement  which has compound listeners can
be expressed as a sum of $\pm$ tanglements whose listeners are smaller(i.e., have fewer nodes).

Note that given a c.m.i. with 
compound listeners, one can: (1) use the 
duality equations to express the c.m.i. as a sum of $\pm$ tanglements;
(2)use the results of this section to express the tanglements 
obtained in step 1 
as a sum of $\pm$ tanglements which have smaller 
listeners;  (3)use the duality equations to express the
tanglements obtained in step 2 as sum of $\pm$ c.m.i.'s.
For example,

\beq
\mu_{1  :  2,3} =
\tau_{1 \tsep 2,3}=
\tau_{1 \tsep 2 \tsep 3} - \tau_{2 \tsep 3}=
\mu_{1:2} + \mu_{1:3} - \mu_{1:2:3}
\;.
\mathHS\eeq
Thus, any c.m.i. which has compound listeners can
expressed as a sum of $\pm$ c.m.i.'s whose listeners are smaller.

Another way of decomposing the compound listeners of a c.m.i. is by using the following 
``chain rule":

\beq
H[\rvX_1 : ( \rvX_2, \rvX_3, \ldots, \rvX_n) | \rvE] =
\sum_{\lam=2}^n H[ \rvX_1 : \rvX_\lam | (\rvX_{\lam + 1}, \ldots, \rvX_n, \rvE) ] 
\;.
\mathHS\eeq
For example,

\beq
H[\rvX_1 : ( \rvX_2, \rvX_3, \rvX_4) | \rvE] =
+H[\rvX_1 : \rvX_2, | \rvX_3, \rvX_4, \rvE] 
+H[\rvX_1 : \rvX_3 | \rvX_4, \rvE]
+H[\rvX_1 : \rvX_4 | \rvE] 
\;.
\mathHS\eeq
This rule is also valid for more than 2 listeners. For example, it can 
be used to decompose the listeners of $\mu((1, 2): (3, 4): (5, 6, 7))$.

\MyCases{(9) Conditionally separable states} 

\begin{figure}[h]
	\begin{center}
	\epsfig{file=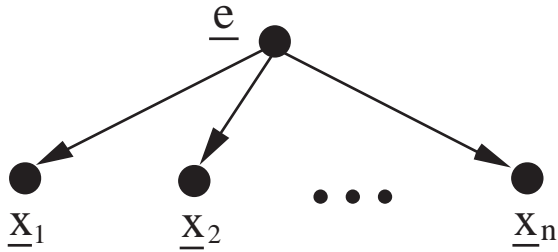, height=1.5in}
	\caption{Net with one speaker and $n$ listener nodes.}
	\label{fig:nlist-pure}
	\end{center}
\end{figure}

Suppose   

\beq
P(X_1, X_2, \ldots, X_n ,  E)  =  P(X_1|E) P(X_2|E)\ldots P(X_n|E) P(E)
\;
\label{eq:n-cond-separ-clas}\eeq
for all values of $X_1, X_2, \ldots, X_n , E$.
Then $HT(\rvX_1\tsep \rvX_2\tsep \ldots \tsep \rvX_n | \rvE) = 0$.
If the speaker $\rvE$ is a single node $\rve$, and for each $\lam$, the listener $\rvX_\lam$ is
a single node $\rvx_\lam$, 
then Eq.(\ref{eq:n-cond-separ-clas}) 
is satisfied by the CB net in Fig.\ref{fig:nlist-pure}.

So far we've only considered the classical case.
The analogous result in the quantum case is as follows.
Suppose that $\rho$ is defined by 

\beq
\rho = \sum_E w_E \ket{E}\bra{E} \rho_E^{(1)} \rho_E^{(2)} \ldots \rho_E^{(n)} 
\;,
\label{eq:n-cond-separ-quant}\eeq
where the $w_E$'s are non-negative numbers that add up to one,
where $\{ \ket{E} | \forall E \}$ is an orthonormal basis for $\hil_\rvE$,
and where for all $\lam\in Z_{1, n}$ and for all $E$, 
$\rho_E^{(\lam)}$ acts on $\hil_{\rvX_\lam}$. The Hilbert spaces 
$\hil_{\rvX_\lam}$ for all $\lam$ and $\hil_\rvE$ are different
spaces. Then $ST_\rho(\rvX_1\tsep \rvX_2\tsep \ldots \tsep \rvX_n | \rvE) = 0$.
If the speaker $\rvE$ is a single node $\rva$, and for each $\lam$, the listener $\rvX_\lam$ is
a single node $\rvx_\lam$, 
then the $\rho$ of Eq.(\ref{eq:n-cond-separ-quant})
can be implemented by a QB net 
with a graph like the one in Fig.\ref{fig:2list-cond-separ},
but such that $\rva$ has $n$ branches instead of just 2.

We showed previously that $\tau_{1 \tsep 2 \tsep  \ldots \tsep  n}=0$
implies $\mu_{1 : 2 : \ldots : n}=0$. The converse statement is
not true (for $n$ larger than 2). Next we will give an example of a situation in which 
the c.m.i. is always zero but the tanglement may be non-zero.

Suppose $n\geq 2$ and $\Gamma_1, \Gamma_2$ are non-empty disjoint sets such that 
$\Gamma_1\cup \Gamma_2 = Z_{1, n}$. In the classical case, 
assume

\beq
P(X_1, X_2, \ldots, X_n ,  E)  =  P[ (X.)_{\Gamma_1} |E] P[ (X.)_{\Gamma_2} |E] P(E)
\;
\eeq
for all values of $X_1, X_2, \ldots, X_n , E$. In the quantum case, assume

\beq
\rho = \sum_E w_E \ket{E}\bra{E} \rho_E^{(1)} \rho_E^{(2)}
\;,
\eeq
where the $w_E$'s are non-negative numbers that add up to one,
and where for $\lam\in Z_{1, 2}$ and for all $E$, 
$\rho_E^{(\lam)}$ acts on $\hil_{(\rvX.)_{\Gamma_\lam}}$.
Then $\mu_{1:2:\ldots:n}=0$. We won't give a completely general 
proof of this theorem. We will only prove it for $n=4$.

One of the duality equations is:

\beq
\mu_{1:2:3:4} = A - B + C
\;,
\mathHS\eeq
where

\beq
A = 
 \tau_{1 \tsep 2}
+\tau_{1 \tsep 3}
+\tau_{1 \tsep 4}
+\tau_{2 \tsep 3}
+\tau_{2 \tsep 4}
+\tau_{3 \tsep 4}
\;,
\mathHS\eeq

\beq
B = 
 \tau_{1 \tsep 2 \tsep 3}
+\tau_{1 \tsep 2 \tsep 4}
+\tau_{1 \tsep 3 \tsep 4}
+\tau_{2 \tsep 3 \tsep 4}
\;,
\mathHS\eeq

\beq
C = \tau_{1 \tsep 2 \tsep 3 \tsep 4} 
\;.
\mathHS\eeq
First suppose that $\Gamma_1 =\{1,2\}$ and $\Gamma_2 =\{3,4\}$.
Then $\tau(\Gamma_1 \tsep \Gamma_2) = 0$. If 
$\Gamma_1'$ (ditto, $\Gamma_2'$) is a non-empty subset of $\Gamma_1$
(ditto, $\Gamma_2$), then, because removing listeners decreases
tanglement,  $\tau(\Gamma_1' \tsep \Gamma_2') = 0$.
Using Eq.(\ref{eq:split-comp-list}) to
decompose the compound listeners of $\tau(\Gamma_1' \tsep \Gamma_2')$, one gets

\beq
\tau(\tsep_{j \in \Gamma_1' \cup \Gamma_2 '} j ) =
\tau(\tsep_{j \in \Gamma_1'} j ) + \tau(\tsep_{j \in \Gamma_2'} j )
\;.
\label{eq:split-2-comp-list}
\mathHS\eeq
Using Eq.(\ref{eq:split-2-comp-list}), one gets

\beq
A = 
 \tau_{1 \tsep 2}
+\tau_{3 \tsep 4}
\;,
\mathHS\eeq

\beq
B = 
2(
 \tau_{1 \tsep 2}
+\tau_{3 \tsep 4}
)
\;,
\mathHS\eeq

\beq
C = \tau_{1 \tsep 2 \tsep 3 \tsep 4} 
\;,
\mathHS\eeq
so 

\beq
\mu_{1:2:3:4} = 
-\tau_{1 \tsep 2}
-\tau_{3 \tsep 4}
+\tau_{1 \tsep 2 \tsep 3 \tsep 4}
=0
\;.
\mathHS\eeq
Next  suppose that $\Gamma_1 =\{1\}$ and $\Gamma_2 =\{2,3,4\}$.
Using Eq.(\ref{eq:split-2-comp-list}), one gets

\beq
A = 
 \tau_{2 \tsep 3}
+\tau_{2 \tsep 4}
+\tau_{3 \tsep 4}
\;,
\mathHS\eeq

\beq
B = 
 \tau_{2 \tsep 3}
+\tau_{2 \tsep 4}
+\tau_{3 \tsep 4}
+\tau_{2 \tsep 3 \tsep 4} 
\;,
\mathHS\eeq

\beq
C = \tau_{2 \tsep 3 \tsep 4}  
\;,
\mathHS\eeq
so 

\beq
\mu_{1:2:3:4}
=0
\;.
\mathHS\eeq

\MyCases{(10) A posteriori local unitary transformations}

In Section \ref{sec:st-2-pure}, we distinguished between a priori and
a posteriori local unitary transformations, and we maximized $ST$ 
over all a priori transformations. Next we will show that $ST$
is in fact invariant under a posteriori local unitary transformation. 
For definiteness, we will calculate $ST$ for a pure state and  2 single-node listeners,
but analogous conclusions hold for a mixed state and $n\geq2$ either single-node or compound listeners.

\begin{figure}[h]
	\begin{center}
	\epsfig{file=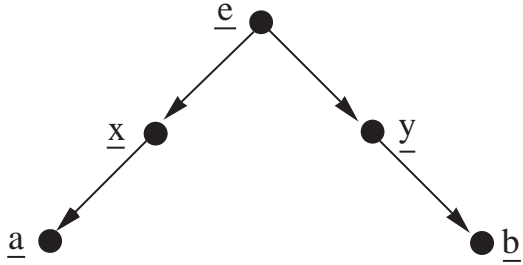, height=1.5in}
	\caption{Net with one speaker node and 2 branches, each branch with 2 nodes.}
	\label{fig:2list-pure-post}
	\end{center}
\end{figure}

Consider the QB net of Fig.\ref{fig:2list-pure-post}, where
\BeginBNetTabular
	$\rve$ & $e = (e_1, e_2)$ & $\psi(e)$ & $\sum_e | \psi(e) | ^2 =1$\\
\hline
	$\rvx$ & $x$ & $\delta(x, e_1)$ & \\
\hline
	$\rvy$ & $y$ & $\delta(y, e_2)$ & \\
\hline
	$\rva$ & $a$ & $U_{ax}$ & $\sum_a U^*_{ax}U_{ax'} = \delta^x_{x'}$\\
\hline
	$\rvb$ & $b$ & $U_{by}$ & $\sum_b U^*_{by}U_{by'} = \delta^y_{y'}$\\
\EndBNetTabular
Let $\qnet$ be the QB net which contains all the nodes shown in Fig.\ref{fig:2list-pure-post}.
Let $\qnet_0$ be the sub-net which contains only nodes $\rve, \rvx$ and $\rvy$. 

The meta density matrix $\mu_0$ of $\qnet_0$ is 

\beq
\mu_0 = \ket{\psi^0_{meta}}\bra{\psi^0_{meta}}
\;,
\eeq
where

\beq
\ket{\psi^0_{meta}} = \sum_{ri} \psi(x,y) \ket{\rve=(x,y), x, y}
\;.
\eeq

The meta density matrix $\mu$ of $\qnet$ is 

\beq
\mu = \rhometa
\;,
\eeq
where

\beq
\ketmeta = \sum_{ri} U_{ax} V_{by}\psi(x,y) \ket{\rve=(x,y), x, y, a, b}
\;.
\eeq
This last equation can be rewritten as

\beq
\ketmeta = \sum_{ri} \psi(x,y) \ket{\rve=(x,y), x, y}\ket{\phi_\rva(x)}\ket{\phi_\rvb(y)}
\;,
\eeq
where

\beq
\ket{\phi_\rva(x)} = \sum_a U_{ax} \ket{a}
\;,\;\;
\ket{\phi_\rvb(y)} = \sum_b V_{by} \ket{b}
\;.
\eeq
The $\ket{\phi_\rva(x)}$'s (ditto, $\ket{\phi_\rvb(y)}$'s )
are an orthonormal basis in $\hil_\rva$ (ditto, $\hil_\rvb$) 
labelled by the indices $x$ (ditto, $y$).

Define $\rho$ by

\beq
\rho = \esum{\rvx, \rvy} (\mu) = 
\sum_{ri} \psi(x,y) \psi^*(x',y') 
\ket{\rve=(x,y), \phi_\rva(x), \phi_\rvb(y)}
\bra{\rve=(x',y'), \phi_\rva(x'), \phi_\rvb(y')}
\;.
\label{eq:rho-def}\eeq
The only difference between $\rho$ and $\mu_0$ is that the
$\phi_\rva(x)$ and $\phi_\rvb(y)$ indices in $\rho$ are replaced by $x$ and $y$ in $\mu_0$.
Thus,

\beq
S_{\rho}(\rva : \rvb | \rve) = S_{\mu_0}(\rvx : \rvy | \rve)
\;.
\eeq
In other words, $ST$ for net $\qnet$, density matrix $\rho$ and listeners $\rva$ and $\rvb$ equals
$ST$ for sub-net $\qnet_0$, density matrix $\mu_0$ and listeners $\rvx$ and $\rvy$. 
Note that in the definition Eq.(\ref{eq:rho-def}) of $\rho$, we e-summed $\mu$ over $\rvx$ and $\rvy$. 
Consider a density matrix $\sigma$ defined by
trace-ing instead of e-summing over $\rvx, \rvy$:

\beq
\sigma = \trace{\rvx, \rvy}(\mu) =
\sum_{ri} \psi(x,y) \psi^*(x,y) 
\ket{\rve=(x,y), \phi_\rva(x), \phi_\rvb(y)}
\bra{\rve=(x,y), \phi_\rva(x), \phi_\rvb(y)}
\;.
\eeq
It is easy to show that

\beq
S_{\sigma}(\rva : \rvb | \rve) = 0
\;.
\eeq
Thus, e-summing over $\rvx$ and $\rvy$ (which corresponds to not measuring those nodes)
gives the same $ST$ as if the local transformations at nodes $\rva, \rvb$ had not occurred.
On the other hand, trace-ing over $\rvx$ and $\rvy$ (which corresponds to measuring those 
nodes in a particular way)
gives zero $ST$, just as in the classical case.

\MyCases{(11) Conditional Data Processing Inequalities}

An introduction to Data Processing (DP) Inequalities for CB and QB nets may be 
found in Ref.\cite{TucciQInfo}.
Here, we will prove a new version of these inequalities
which we call Conditional DP Inequalities. The Conditional DP 
Inequalities are conditioned on a speaker.
Thus, they are closely linked to the phenomenon of tanglement.
Consider the net of Fig.\ref{fig:2list-pure-post}. What we will show is that 

\beq
H(\rva : \rvb | \rve) \leq H(\rvx : \rvy | \rve)
\;.
\label{eq-cond-dp-ineq}
\mathHS\eeq

In the quantum case, we've shown  in the 
previous section entitled ``A posteriori local unitary transformations"
that if nodes $\rva$ and $\rvb$ correspond to unitary transformations and 
nodes $\rvx$ and $\rvy$ to delta functions, then 
equality is attained in inequality Eq.(\ref{eq-cond-dp-ineq}). No such assumptions
about the nature of the transition matrices of the nodes will be made in this 
section. Our assumptions are only that the QB net has a particular 
topology, that of Fig.\ref{fig:2list-pure-post}. 

Clearly, the Conditional DP Inequalities of this section can be 
greatly generalized in the same way that Ref.\cite{dp-ineq} generalizes 
DP Inequalities from a simple Markov chain to arbitrary CB or QB nets. 
In this section, we will discuss only the 
simplest case of the Conditional DP Inequalities. 
More general cases will be discussed in a future paper
dedicated exclusively to this subject.

Eq.(\ref{eq-cond-dp-ineq}) has a simple interpretation, as all DP inequalities do.
It says that the conditional information transmission between $\rvx$ and
$\rvy$ is larger than that between $\rva$ and
$\rvb$ because the first pair of nodes is ``closer". 
Alternatively, one can say that the probabilistic 
dependency of $\rvx$ on
$\rvy$ is larger than that between $\rva$ and
$\rvb$ because the first pair of nodes is ``closer".

First note that the graph of Fig.\ref{fig:2list-pure-post} satisfies

\beq
H(\rva | \rve, \rvy, \rvb) = H(\rva | \rve, \rvy)
\;.
\label{eq:indep-of-b}
\mathHS\eeq
In the classical case, this follows because 
$P(\rva | \rve, \rvy, \rvb) = P(\rva | \rve, \rvy)$.
By virtue of Eq.(\ref{eq:indep-of-b}) and strong subadditivity,

\beq
H(\rva | \rve, \rvy) = H(\rva | \rve, \rvy, \rvb) \leq H(\rva | \rve, \rvb)
\;.
\mathHS\eeq
Subtracting $H(\rva | \rve)$ from each term of the last equation and multiplying
the resulting equation by $-1$ gives

\beq
H(\rva : \rvy | \rve) \geq
H(\rva : \rvb | \rve)
\;.
\label{eq:fixed-a}\mathHS\eeq

Now note that the graph of Fig.\ref{fig:2list-pure-post} satisfies

\beq
H(\rvy | \rve, \rvx, \rva) = H(\rvy | \rve, \rvx)
\;.
\label{eq:indep-of-a}
\mathHS\eeq
In the classical case, this follows because 
$P(\rvy | \rve, \rvx, \rva) = P(\rvy | \rve, \rvx)$.
By virtue of Eq.(\ref{eq:indep-of-a}) and strong subadditivity,

\beq
H(\rvy | \rve, \rvx) = H(\rvy | \rve, \rvx, \rva) \leq H(\rvy | \rve, \rva)
\;.
\mathHS\eeq
Subtracting $H(\rvy | \rve)$ from each term of the last equation and multiplying
the resulting equation by $-1$ gives

\beq
H(\rvy : \rvx | \rve) \geq
H(\rvy : \rva | \rve)
\;.
\label{eq:fixed-y}
\mathHS\eeq

Combining Eqs.(\ref{eq:fixed-a}) and (\ref{eq:fixed-y}) gives

\beq
H(\rva : \rvb | \rve) \leq
H(\rva : \rvy | \rve) \leq
H(\rvx : \rvy | \rve)
\;.
\mathHS\eeq
QED.

\begin{figure}[h]
	\begin{center}
	\epsfig{file=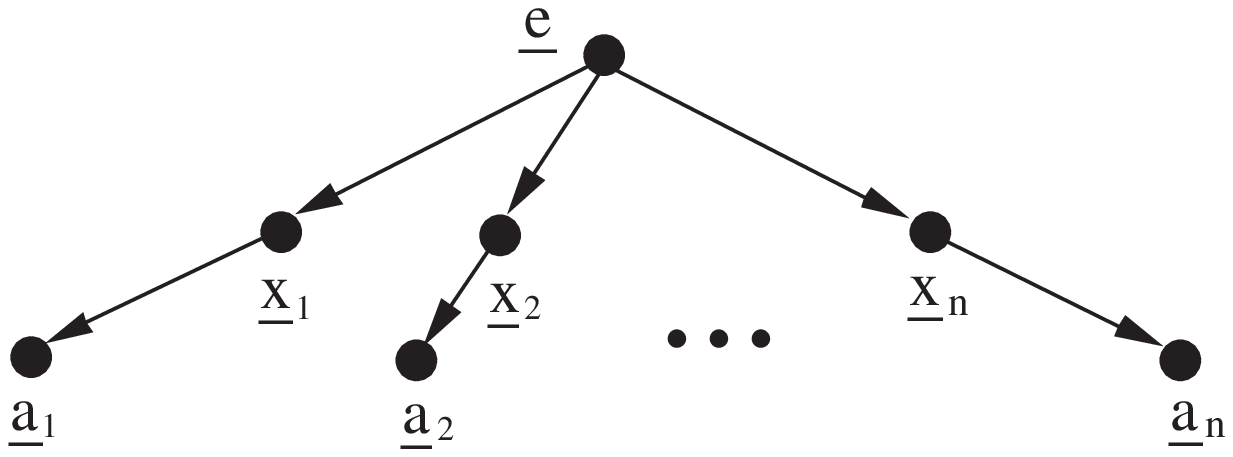}
	\caption{Net with one speaker node and $n$ branches, each branch with 2 nodes.}
	\label{fig:nlist-pure-post}
	\end{center}
\end{figure}

Eq.(\ref{eq-cond-dp-ineq}) can be easily generalized to $n\geq 2$ listeners. 
Consider the graph of Fig.\ref{fig:nlist-pure-post}.
Next  we will show that for this graph, 

\beq
HT(\rva_1 \tsep \rva_2 \tsep \ldots \tsep \rva_n | \rve) \leq HT(\rvx_1 \tsep \rvx_2 \tsep \ldots \tsep \rvx_n | \rve)
\;.
\label{eq-cond-dp-ineq-n}\mathHS\eeq
The proof is by induction on $n\geq 2$. Eq. (\ref{eq-cond-dp-ineq-n}) has 
been proven for $n=2$. If  it is true
for all $n\in Z_{2, n_0}$, then is must be true for $n=n_0+1$. Here is why.
By virtue of the induction hypothesis, the following two inequalities must be
true:

\beq
HT[ (\rva_1 , \rva_2 , \ldots , \rva_{n_0}) \tsep\rva_{n_0+1} | \rve]
\leq
HT[ (\rvx_1 , \rvx_2 , \ldots , \rvx_{n_0}) \tsep \rvx_{n_0+1} | \rve]
\;,
\mathHS\eeq

\beq
HT[ \rva_1 \tsep \rva_2 \tsep \ldots \tsep \rva_{n_0} | \rve]
\leq
HT[ \rvx_1 \tsep \rvx_2 \tsep \ldots \tsep \rvx_{n_0} | \rve]
\;.
\mathHS\eeq
The sum of the left-hand sides (ditto, right-hand sides) of these two inequalities equals
$HT(\rva_1 \tsep \rva_2 \tsep \ldots \tsep \rva_{n_0+1} | \rve)$
(ditto, $HT(\rvx_1 \tsep \rvx_2 \tsep \ldots \tsep \rvx_{n_0+1} | \rve)$) \HS. QED

\begin{appendix}
\BeginSection{Proof that $E_F(|\psi_{xy}|)\leq E_F(\psi_{xy})$ }\label{app:ef-ineq}

We will first prove this inequality for the case that $S_\rvx = S_\rvy = Bool$.
Define the function $p_0(t)$ for $t\in [0,1]$ by 

\beq
p_0(t) = \frac{1 + \sqrt{ 1 - t}}{2}
\;.
\eeq
From Eqs.(\ref{eq:ef-px}) and (\ref{eq:p0-t-def}), 

\beq
E_F(\psi_{xy}) = h(p_0(t))
\;,
\eeq
where

\beq
t =  4 | \psi_{00} \psi_{11} - \psi_{01}\psi_{10} |^2
\;.
\eeq
Let 

\beq
t' = 4 ( |\psi_{00} \psi_{11}| - |\psi_{01}\psi_{10} |)^2
\;.
\eeq
Note that

\beq
E_F( |\psi_{xy}|) = h(p_0(t'))
\;.
\eeq
By the triangle inequality, 

\beq
t' \leq t
\;.
\eeq
From Fig.\ref{fig:po-h-plot}, $h(p_0(t))$ is a monotonically increasing function of $t$. Thus

\beq
E_F( |\psi_{xy}|) = h(p_0(t')) \leq h(p_0(t)) = E_F( \psi_{xy})
\;.
\eeq

Now consider the case of arbitrary $N_\rvx, N_\rvy$ such that $N_\rvx \leq N_\rvy$. Recall

\beq
E_F(\psi_{xy}) = S(\rho)
\;,
\eeq
where

\beq
\rho = \psi \psi^\dagger
\;.
\eeq
For all $x, y$, define $\theta_{xy}$ to be the phase of $\psi_{xy}$.
Then

\beq
\rho_{xx'} =
\sum_y \psi_{xy} \psi^*_{x'y}
=
\sum_y e^{i (\theta_{xy} - \theta_{x'y})} |\psi_{xy}\psi_{x'y}|
\;.
\eeq
Suppose we vary the angles $\theta_{xy}$. Then

\beq
\delta S(\rho) =
-\delta \tr \left[ \rho \frac{\ln \rho}{\ln 2} \right] =
-\tr\left[ \frac{\delta \rho}{\ln 2} ( \ln \rho + 1) \right]
\;,
\eeq
where

\beq
\delta \rho_{xx'} =
\sum_y i(\delta \theta_{xy} - \delta \theta_{x'y}) \psi_{xy} \psi^*_{x'y}
\;.
\eeq
When $\theta_{xy} = 0$ for all $x$ and $y$, 
$\delta \rho_{xx'}$ is antisymmetric
and 
$\rho_{xx'}$ is symmetric 
under the exchange of $x$ and $x'$. If $A$ and $S$ are, respectively, 
an antisymmetric and a symmetric $N\times N$ matrix, then
$\tr(A) = \tr(AS)=0$. Thus, $\tr(\delta \rho)=\tr(\delta \rho \ln \rho)=0$.
Thus, $\delta S(\rho) = 0$ when $\theta_{xy} = 0$ for all $x$ and $y$.
I don't know how to show for general values of $N_\rvx$ and $N_\rvy$
that this extremum of $S(\rho)$ is a global minimum.

\BeginSection{Proof that $H(\rvx : \rvy) \leq E_F(\sqrt{P_{xy}})$ }\label{app:mut-info-bound}

In this appendix, we will prove an inequality which gives
an upper bound for the classical mutual information $H(\rvx : \rvy)$.
From $H(\rvx : \rvy) = H(\rvx) - H(\rvx | \rvy)$ 
and $H(\rvx | \rvy)\geq 0$, it follows that

\beq
H(\rvx : \rvy) \leq \min\{ H(\rvx), H(\rvy) \}
\;.
\eeq
What we seek here is a tighter upper bound for $H(\rvx : \rvy)$.

Suppose $\rvx$ (ditto, $\rvy$) is a random variable that can assume
values in a set $S_\rvx$ (ditto, $S_\rvy$) which contains $N_\rvx$ 
(ditto, $N_\rvy$) elements. Let $P_{xy}$ be the joint probability 
distribution of $\rvx$ and $\rvy$. Let $P_{x-} = \sum_y P_{xy}$
and $P_{-y} = \sum_x P_{xy}$. Without loss of generality,
we will assume that $N_\rvx \leq N_\rvy$.

Define $\Psi$ to be the rectangular matrix with entries

\beq
\Psi_{xy} = \sqrt{P_{xy}}
\;.
\eeq
 Note that

\beq
\tr(\Psi \Psi^T) = \sum_{x,y} \Psi_{xy}^2 = \sum_{x,y} P_{xy} = 1
\;.
\eeq
Let

\beq
\markPsi = U \Psi V^T
\;,
\eeq
where $U$ and $V$ are (real) orthogonal matrices. Define 

\beq
\markP_{xy} = \markPsi_{xy}^2
\;.
\eeq
Then

\beq
\sum_{x,y} \markP_{xy} = \tr(\markPsi \markPsi^T) = \tr(\Psi \Psi^T)  = 1
\;.
\eeq

Define $\eta$ by

\beq
\eta = \sum_{x,y} \markP_{xy} \ln \frac{ \markP_{xy}}{ \markP_{x-} \markP_{-y}}
\;.
\eeq
Note that

\beq
H(\rvx : \rvy) = \left. \frac{\eta}{\ln 2} \right|_{U=V=1}
\;,
\eeq
where the right-hand side is evaluated at $U=V=1$.
Our goal is to show that: (1) $\eta$ has a global maximum 
when it varies over the spaces of all orthogonal $N_\rvx \times N_\rvx$
matrices $U$ and all orthogonal $N_\rvy \times N_\rvy$ matrices $V$; (2) the maximum 
occurs when 
$U$ and $V$ make $\markPsi$ diagonal. (Such a 
$U$ and $V$ exist by the Singular Value Decomposition Theorem).
When $\markPsi$ is diagonal, 

\beq
\frac{\eta}{\ln 2} =
\sum_x \markP_{xx} \log_2 \frac{1}{ \markP_{xx} }
=E_F(\markPsi_{xy})
=E_F(\Psi_{xy}) 
=E_F(\sqrt{P_{xy}}) 
\;.
\eeq
Therefore, if $\eta$ has a global maximum when $\markPsi$ is diagonal,
then

\beq
H(\rvx : \rvy) \leq E_F(\sqrt{P_{xy}})
\;.
\eeq

Suppose we vary each $\markP_{xy}$ by $\delta\markP_{xy}$
in such a way that 

\beq
\sum_{x,y} \delta\markP_{xy} = 0
\;.
\label{eq:zero-delta-sum}\eeq
(And therefore also $\sum_x \delta\markP_{x-} = \sum_y \delta\markP_{-y} = 0$.) Then

\beq
\delta \eta = 
\sum_{x,y} (\delta\markP_{xy}) \ln \frac{ \markP_{xy}}{ \markP_{x-} \markP_{-y}} 
+\;\;nil
\;,
\label{eq:delta-eta}\eeq
where 

\beq
nil = 
\sum_{x,y} 
\left(
\delta\markP_{xy}
- \frac{\markP_{xy}}{\markP_{x-}} \delta\markP_{x-}
- \frac{\markP_{xy}}{\markP_{-y}} \delta\markP_{-y}
\right)
\;
\eeq
Because of Eq.(\ref{eq:zero-delta-sum}), $nil = 0$.

$U$ and $V$ are orthogonal and we will vary them so that 
$U + \delta U$ and $V + \delta V$ are also orthogonal. Thus,
$\sum_{xy} (\markP_{xy} + \delta \markP_{xy}) = 1$. Thus,
Eq.(\ref{eq:zero-delta-sum}) is satisfied. 

For $N_\rvx=N_\rvy=2$, $U$ and $V$ can be parameterized by expressing them as

\beq
U=
\left[
\begin{array}{cc}
c_1 & s_1 \\
-s_1 & c_1
\end{array}
\right ]
\;,\;\;
V=
\left[
\begin{array}{cc}
c_2 & s_2 \\
-s_2 & c_2
\end{array}
\right ]
\;,
\eeq
where $c_j =\cos \theta_j$, $s_j =\sin \theta_j$ for $j=1,2$. 
Then we can vary $U$ and $V$ by varying the angles $\theta_1, \theta_2$.
For general $N_\rvx$ and  $N_\rvy$,  we can express $U$ and $V$ as $U=e^\alpha$ and
$V = e^\beta$, where $\alpha$ and $\beta$ are antisymmetric matrices.
Then we can vary $U$ and $V$ by varying 
the components of $\alpha$ and $\beta$ 
that lie above their main diagonal.

One gets

\beq
\delta \markP_{xy} = 2 \markPsi_{xy} \delta \markPsi_{xy}
\;,
\label{eq:delta-mark-p}\eeq
and

\beq
\delta \markPsi = 
(\delta U) \Psi V^T
+ U \Psi (\delta V^T)
=
A \markPsi
+ \markPsi B
\;,
\label{eq:delta-mark-psi}\eeq
where

\beq
A= (\delta U) U^T, \;\; B= V \delta V^T
\;.
\eeq
Because $UU^T=1$, $(\delta U) U^T + U \delta U^T = 0$,  
which can be expressed in terms of $A$ as $A = -A^T$,
Thus, $A$ must be antisymmetric.
$B$ must be antisymmetric too.

Next we will show that if $U$ and $V$ 
are such that $\markPsi$ is diagonal, then $\delta\markP_{xy} = 0$ for all $x$ and $y$,
and therefore, by Eq.(\ref{eq:delta-eta}), $\delta \eta = 0$.
Consider some $x, y$ such that $x\neq y$; for example, 
$x=0, y=1$. 
Since $\markPsi_{01}=0$, Eq.(\ref{eq:delta-mark-p}) implies
$\delta\markP_{01} = 0$. 
Consider some $x, y$ such that $x=y$; for example, 
$x=y=0$. $\sum_a A_{0a} \markPsi_{a0}= 0$
because when $a=0$, $A_{00}=0$, and when $a\neq 0$, 
$\markPsi_{a0}=0$. Likewise,  $\sum_b \markPsi_{0b} B_{b0} = 0$.
Thus, by Eq.(\ref{eq:delta-mark-psi}), $\delta\markPsi_{00} = 0$.
Since $\delta\markPsi_{00}=0$, Eq.(\ref{eq:delta-mark-p}) implies
$\delta\markP_{00} = 0$.

So far we have shown that $\delta \eta =0$ when $\markPsi$ is diagonal. It
remains for us to show that this extremum is a global maximum.
I don't know how to show this. However,
my Monte Carlo tests support this claim. Furthermore, the following
argument shows that the extremum is at least a local maximum.
One has 

\beq
\delta^2 \eta = 
\sum_{x,y} (\delta^2 \markP_{xy}) \ln \left( \frac{ \markP_{xy} } { \markP_{x-} \markP_{-y} } \right)
+ nil'
\;,
\eeq
where

\beq
nil' = 
\sum_{x,y} \frac{ (\delta \markP_{xy})^2 }{ \markP_{xy} }
-\sum_{x} \frac{ (\delta \markP_{x-})^2 }{ \markP_{x-} }
-\sum_{y} \frac{ (\delta \markP_{-y})^2 }{ \markP_{-y} }
\;.
\eeq
If $\markPsi$ is diagonal, then $\delta\markP_{xy} = 0$ for all $x$ and $y$
so $nil'=0$. One has

\beq
\delta^2 \markP_{xy} 
=
\delta[ 2 \markPsi_{xy} \delta \markPsi_{xy}] =
2 (\delta \markPsi_{xy})^2 + 2 \markPsi_{xy} \delta^2 \markPsi_{xy}
\;.
\eeq
If $\markPsi$ is diagonal, then 
$\delta^2 \markP_{xy} = 2 (\delta \markPsi_{xy})^2 \geq 0$ for any
$x\neq y$. But $\markP_{xy}=0$ for $x\neq y$ so
$\delta^2\eta \rarrow -\infty$. Thus $\eta$
has a local maximum when $\markPsi$ is diagonal.
In fact, $\eta$ has a cusp there. The cusp is on
the boundary of the region on which 
$\markP_{xy}$ is defined.

\BeginSection{Entropy of Density Matrix \\with Repeated Index Pairs}\label{app:rep-indices}
Often in this paper we need to evaluate the entropy of a density matrix
such as 

\beq
R = \sum_{a, a'} R_{a, a'} \ket{\rva = a, \rvb = a} \bra{\rva = a', \rvb = a'} 
\;,
\eeq
where the nodes $\rva$ and $\rvb$ have the same states ($S_\rva = S_\rvb$).
By an ``index pair" of a matrix $M$ we mean the row and column indices of an entry of $M$.
The index pair $(a,a')$ is repeated in $R$.
Consider the smaller density matrix

\beq
\rho = \sum_{a, a'} R_{a, a'} \ket{a} \bra{a'}
\;.
\eeq
Next we will show that $S(R) = S(\rho)$. Thus, for the 
purpose of evaluating its entropy, one can replace the density
matrix $R$ by the smaller $\rho$.  The proof consists of 
showing that $R$ and $\rho$ have the same non-zero eigenvalues.
Indeed, suppose $\ket{\phi}\in \hil_\rva$ is an eigenvector
of $\rho$: 

\beq
\rho \ket{\phi} = \lambda \ket{\phi}
\;.
\eeq
Then  $\ket{\Phi}$ defined by

\beq
\ket{\Phi} = \sum_{a'} \ket{\rva=a', \rvb=a'} \av{\rva=a'|\phi}
\;
\eeq
is an eigenvector of $R$ with the same eigenvalue $\lambda$. Indeed,

\beq
R\ket{\Phi} = \sum_{a, a'} R_{a, a'} \ket{\rva = a, \rvb = a} \av{\rva=a'|\phi} 
= \lambda\ket{\Phi} 
\;.
\eeq
Thus, the set of eigenvalues of $R$ contains the set of eigenvalues of $\rho$. From the matrix
representation of $R$, it is clear that any eigenvalue of $R$ 
which is not an eigenvalue of $\rho$ must be zero.

\end{appendix}


\begin{thebibliography}{99}
%
\bibitem{NoHootters}C.H. Bennett, H.J. Bernstein, S. Popescu, B. Schumacher, 
Phys. Rev. A {\bf 53} (1996) 2046. Also available as Los Alamos eprint
quant-ph/9511030.

\bibitem{LittleHootters}C.H. Bennett, G. Brassard, S. Popescu, B. Schumacher, J. Smolin, W.K. Wootters,
Phys. Rev. Lett. {\bf 76} (1996) 722. Also available as Los Alamos eprint quant-ph/9511027.

\bibitem{MonsterHootters}
C.H. Bennett, D.P. DiVincenzo, J.A. Smolin, W.K. Wootters,
Phys. Rev. A {\bf 54} (1996) 3824-3851. Also available as Los Alamos eprint
quant-ph/9604024.

\bibitem{mixture}Here is a small sampling of entanglement papers
published at the Los Alamos
eprint library in just the last 2  months! :
%
V. Coffman, J. Kundu, W.K. Wootters,
quant-ph/9907047 ;
%
M. Horodecki, P. Horodecki, R. Horodecki, 
quant-ph/9908065 ;
%
C. Bennett, S. Popescu, D. Rohrlich, J. Smolin, A.V. Thaphiyal, 
quant-ph/9908073 ;
%
L. Henderson, V. Vedral,
quant-ph/9909011 ;

\bibitem{TucciQInfo}
R.R. Tucci, ``Quantum Information Theory - A Quantum Bayesian Net Perspective", Los Alamos eprint quant-ph/9909039.

\bibitem{QFog}See, for example, 
``Quantum Fog Library of Essays", which can be downloaded for free at
www.ar-tiste.com 

\bibitem{Hootters93}
L.P. Hughston, R. Jozsa, W.K. Wootters, Phys. Lett. A {\bf 183} (1993) 14-18.

\bibitem{Noble}B. Noble and J.W. Daniels, 
{\it Applied Linear Algebra}, Third Edition (Prentice Hall, 1988).

\bibitem{McGill}
W.J. McGill, 
``Multivariate Information Transmission", IRE Trans. Info. Theory {\bf 4} (1954) 93-111.

\bibitem{dp-ineq}
R.R. Tucci, ``Data Processing Inequalities for Bayesian Nets", Los Alamos eprint quant-ph/?

\end{thebibliography}
\end{document}